\documentclass[aps,pra,twocolumn]{revtex4-1}

\usepackage{latexsym}
\usepackage{float}

\usepackage{mathrsfs}
\usepackage{graphicx,amsfonts,amssymb}
\usepackage[fleqn]{amsmath}
\usepackage{epstopdf}
\usepackage{hyperref}

\begin{document}

% Use the \preprint command to place your local institutional report
% number in the upper righthand corner of the title page in preprint mode.
% Multiple \preprint commands are allowed.
% Use the 'preprintnumbers' class option to override journal defaults
% to display numbers if necessary
%\preprint{}

%Title of paper
\title{BCS pairing state of a Dilute Bose Gas with
Spin-Orbit Coupling}

% repeat the \author .. \affiliation  etc. as needed
% \email, \thanks, \homepage, \altaffiliation all apply to the current
% author. Explanatory text should go in the []'s, actual e-mail
% address or url should go in the {}'s for \email and \homepage.
% Please use the appropriate macro foreach each type of information

% \affiliation command applies to all authors since the last
% \affiliation command. The \affiliation command should follow the
% other information
% \affiliation can be followed by \email, \homepage, \thanks as well.

\author{Dekun Luo, Lan Yin}
\email[]{yinlan@pku.edu.cn}

%\homepage[]{Your web page}
%\thanks{}
%\altaffiliation{}

\affiliation{School of Physics, Peking University, Beijing 100871, China}

%Collaboration name if desired (requires use of superscriptaddress
%option in \documentclass). \noaffiliation is required (may also be
%used with the \author command).
%\collaboration can be followed by \email, \homepage, \thanks as well.
%\collaboration{}
%\noaffiliation

\date{\today}

\begin{abstract}
We study a two-component Bose gas with a symmetric spin-orbit coupling, and find that two atoms can form a bound state with any intra- or inter-species scattering length.  Consequently, in the dilute limit, the Bardeen-Cooper-Shrieffer (BCS) pairing state of bosons can be formed with weakly-attractive inter-species and repulsive intra-species interactions.   The quasiparticle excitation energies are anisotropic due to spin-orbit coupling.  This BCS paring state is energetically favored over Bose-Einstein condensation (BEC) of atoms at low densities.  As the density increases, there is a first-order transition from the BCS to BEC states.
\end{abstract}

% insert suggested PACS numbers in braces on next line
\pacs{}
% insert suggested keywords - APS authors don't need to do this
%\keywords{}

%\maketitle must follow title, authors, abstract, \pacs, and \keywords

\maketitle
\section{Introduction}
Observation of BCS-BEC crossover in Fermi gases was a tremendous triumph in the research of ultracold quantum gases \cite{*[{For detailed introduction,see }] 010101}.  In contrast, although it was proposed around half a century ago\cite{ISI:A1958WQ70400002,ISI:A1969E277200015}, the BCS state of bosons has never been observed.  The BCS state of a Bose gas with Feshbach resonance was theoretically studied \cite{intro1,intro2,intro3}, and it was found that this state is generally unstable in the attractive region or close to the resonance \cite{intro3,intro4,intro5,intro6}. In experiments, the lifetime of Feshbach molcules was too short for equilibrating into a BEC state \cite{010401,010402,010403,010404}.  Here we propose that the BCS state with a Bose gas can be realized in a Bose with a three-dimensional (3D) spin-orbit coupling (SOC).

The SOC of ultracold atomic gases was experimentally realized in Bose gases \cite{intro20102,intro20101} and Fermi gases \cite{intro20201,intro20202}.  In contrast to SOC of electrons, the SOC of ultracold atoms refers to the coupling between spin of the atomic internal state and momentum of the atom \cite{intro20304,intro20305,intro20307,intro20310}.  The experimental realization of
SOC in cold atoms provides a new platform for studying spin-orbit-coupled many-body systems \cite{intro20304j01,intro20304j02,intro20304,intro20304jg03}.  It can provide simulations of complex phenomena, such as the quantum spin Hall effect\cite{intro20403,intro20305}, topological insulators and superconductors \cite{intro205g01,intro20502}, Majorana fermions\cite{intro20603} and spintronics \cite{intro20701}. So far most of experimental SOC was one-dimensional (1D), and more recently two-dimensional (2D) SOC was realized experimentally \cite{020101,020102}.  Many theoretical work have been focused on phase diagrams of Bose gases with 1D and 2D SOC \cite{intro21701,intro21702,intro21703,intro21503,intro21504,intro21306,intro21402,intro21314jg01}.   There have been proposals to generate 3D SOC \cite{020201,020202} in a Bose gas which is under theoretical investigation \cite{intro21405,intro22001,PhysRevLett.115.253902}.

The realization of SOC in cold atoms may offer the opportunity to realize the long-sought BCS pairing state of Bose atoms.
A pairing condensation in a dilute Bose gas with 2D Rashba SOC and weak intra-species attraction can be stablized by inter-species repulsion \cite{intro301}, but the intra-species attraction can also lead to phase separation which may become an experimental obstacle.  In this work, we investigate the pairing state of Bose gas with an isotropic 3D SOC.  First, we study the two-body bound state of Bose atoms with 3D SOC and find that the bound state can exist for arbitrary inter-species and intra-species scattering length, which is helpful in forming a BCS pairing state.  Next, we study the molecular condensation in a dilute Bose gas with 3D SOC in the framework of the BCS theory.  We find that this pairing state can be stable in the case with weak inter-species attraction and intra-species repulsion which avoids phase separation.  The quasi-particle excitation energy is anisotropic due to SOC.  As the atomic density increases, there is a first-order phase transition from the BCS pairing state to the atomic BEC.  We discuss the experimental perspective of realizing the BCS pairing state of bosons and conclude in the end.

\section{Two-boson bound state with SOC}
\subsection{Model}
We study a two-component homogeneous Bose gas described by the Hamiltonian $H=H_{0}+H_{int}$,
where the single-particle Hamiltonian is given by
\begin{equation}
H_0=\sum_{\textbf{k},\rho,\rho'}c^{\dag}_{\textbf{k}\rho}[\epsilon_{\textbf{k}}  \delta_{\rho\rho'}+\frac{\hbar^{2}\kappa}{m} \textbf{k}\cdot {\bf \sigma}_{\rho\rho'}] c_{\textbf{k}\rho'},
\end{equation}
and the interaction between atoms is given by
\begin{eqnarray}
H_{int}=\frac{1}{2V}\sum_{\textbf{kk}'\textbf{q}\rho\rho'}g_{\rho\rho'}c^{\dag}_{\frac{\textbf{q}}{2}+\textbf{k}'\rho}
c^{\dag}_{\frac{\textbf{q}}{2}-\textbf{k}'\rho'}c_{\frac{\textbf{q}}{2}-\textbf{k}\rho'}c_{\frac{\textbf{q}}{2}+\textbf{k}\rho}.
\end{eqnarray}
Here ${\bf \sigma}_{\rho\rho'}$ are Pauli matrices, $m$ is the atomic mass, $c_{\textbf{k}\rho}$ is the annihilation operator of a Boson with wavevector $\textbf{k}$ and spin component $\rho$, $\epsilon_{\textbf{k}}=\hbar^2k^2/2m$, $\kappa$ is the strength of isotropic 3D SOC, V is the volume, $g_{\uparrow\uparrow}=g_{\downarrow\downarrow}$ is the intra-species coupling constant, and $g_{\uparrow\downarrow}=g_{\downarrow\uparrow}$ is the inter-species coupling constant. The single-particle Hamiltonian can be easily diagonalized, yielding two helicity branches of atomic excitations with eigenenergies $\epsilon_k\pm\hbar^{2}\kappa k/m$.

\subsection{Two-body bound state}
The wavefunction of a two-body bound state satisfies the eigenequation $H|\phi\rangle=E_{\textbf{q}}|\phi\rangle$, where $\hbar\textbf{q}$ is the center of mass momentum, and $E_{\textbf{q}}$ is the eigenenergy.  It can be generally written as
\begin{eqnarray}
|\phi\rangle=\sum_{\textbf{k}\rho\rho'}\!'\psi_{\rho\rho'}(\textbf{k},\textbf{q}-\textbf{k})c^{\dag}_{\textbf{k}\rho}
c^{\dag}_{\textbf{q}-\textbf{k}\rho'}|0\rangle.
\end{eqnarray}
Due to Bose statistics, the coefficients satisfy the symmetric condition $\psi_{\rho\rho'}(\textbf{k},\textbf{k}')=\psi_{\rho'\rho}(\textbf{k}',\textbf{k})$.
From the eigenequation, we obtain the following matrix equation for the coefficients at $\textbf{q}=0$,
\begin{eqnarray}
M_{\textbf{k}}\psi'_{\textbf{k}}=\frac{1}{V}G\sum_{\textbf{p}}\psi'_{\textbf{p}},
\end{eqnarray}
where $\psi'_{\textbf{k}}$ is a four-component vector given by $\psi'_{\textbf{k}}
=[\psi_{\uparrow\uparrow}(\textbf{k},-\textbf{k}),\psi_{\downarrow\downarrow}(\textbf{k},-\textbf{k}),
\psi_{\uparrow\downarrow}(\textbf{k},-\textbf{k}),\psi_{\uparrow\downarrow}(-\textbf{k},\textbf{k})]$,
and $G$ is the matrix of coupling constants
\begin{eqnarray}
G=\left(
\begin{matrix}
g_{\uparrow\uparrow}&0&0&0\\
0&g_{\downarrow\downarrow}&0&0\\
0&0&g_{\uparrow\downarrow}&0\\
0&0&0&g_{\uparrow\downarrow}
\end{matrix}
\right).
\end{eqnarray}
The matrix $M_{\textbf{k}}$ is given by
\begin{eqnarray}
M_{\textbf{k}}=\left(
\begin{matrix}
\varepsilon_{\textbf{k}}&0&S^{*}(\textbf{k}_{\perp})&-S^{*}(\textbf{k}_{\perp})\\
0&\varepsilon_{\textbf{k}}&-S(\textbf{k}_{\perp})&S(\textbf{k}_{\perp})
\\
S(\textbf{k}_{\perp})&-S^{*}(\textbf{k}_{\perp})&\varepsilon_{\textbf{k}}-\frac{2\hbar^{2}\kappa k_{z}}{m}&0
\\
-S(\textbf{k}_{\perp})&S^{*}(\textbf{k}_{\perp})&0&\varepsilon_{\textbf{k}}+\frac{2\hbar^{2}\kappa k_{z}}{m}
\end{matrix}
\right),\nonumber\\
\end{eqnarray}
where $\varepsilon_{\textbf{k}}=E_{0}-2\epsilon_{\textbf{k}}$, $\textbf{k}_{\perp}$ is the projection of $\textbf{k}$ in the $x-y$ plane, and $S(\textbf{k}_{\perp})=\hbar^{2}\kappa(k_{x}+ik_{y})/m$.
Define a new vector
\begin{eqnarray}
Q=\frac{1}{V}G\sum_{\textbf{k}}\psi'_{\textbf{k}},
\end{eqnarray}
and from the eigenequation we obtain
\begin{eqnarray}\label{BDEQ}
Q=\frac{1}{V}G\sum_{\textbf{k}}M_{\textbf{k}}^{-1}Q.
\end{eqnarray}
The sum of matrix $M_{\textbf{k}}^{-1}$ has the following explicit form
\begin{eqnarray*}
 \sum_{\textbf{k}}M_{\textbf{k}}^{-1}=\sum_{\textbf{k}}\frac{(E_{0}m-\hbar^{2}k^{2})}{m^{4}}\det|M_{\textbf{k}}^{-1}|
\left[
\begin{matrix}
a_{\textbf{k}}&0&0&0\\
0&a_{\textbf{k}}&0&0\\
0&0&b_{\textbf{k}}&d_{\textbf{k}}\\
0&0&d_{\textbf{k}}&b_{\textbf{k}}
\end{matrix}
\right],
\end{eqnarray*}
where
\begin{eqnarray*}
 &&a_{\textbf{k}}=m[E_{0}^{2}m^{2}-2E_{0}\hbar^{2}m k^2+\hbar^{4}(k^4-2\kappa^{2}(k^2+k_{z}^{2}))],\\
 &&b_{\textbf{k}}=m[E_{0}^{2}m^{2}-2E_{0}\hbar^{2}m k^2+\hbar^{4}(k^4-2\kappa^{2}k_{\perp}^2)],\\
 &&d_{\textbf{k}}=-2\hbar^{4}m\kappa^{2}k_{\perp}^{2}.
\end{eqnarray*}
For $Q=[q_{1},q_{2},q_{3},q_{4}]$, equation (\ref{BDEQ}) leads to three physical solutions:
(a) $q_{1}\neq0$, $q_{2}=q_{3}=q_{4}=0$; (b) $q_{2}\neq0$, $q_{1}=q_{3}=q_{4}=0$; (c) $q_{1}=q_{2}=0$, $q_{3}=q_{4}\neq0$.
The first two solutions are due to intra-species interaction, and the last solution is due to inter-species interaction.
The unphysical solution with $q_3 \neq q_4$ can be neglected due to symmetry.  Eigenenergies of these bound states satisfy the following equation
\begin{eqnarray}
\frac{m}{4\pi\hbar^{2}a_{\rho\rho'}}=&&\frac{1}{2V}\sum_{\textbf{k}}[\frac{1}{\epsilon_{\textbf{k}}}+\frac{2}{E_{0}-2\epsilon_{\textbf{k}}}\nonumber\\
&&+\frac{16\epsilon_{\textbf{k}_{\perp}}\epsilon_{\kappa}}
{(E_{0}-2\epsilon_{\textbf{k}})^{3}-16\epsilon_{\textbf{k}}\epsilon_{\kappa}(E_{0}-2\epsilon_{\textbf{k}})}], \label{bound2}
\end{eqnarray}
where $a_{\rho\rho'}$ is the scattering length.

\begin{figure}
\includegraphics[width=8cm]{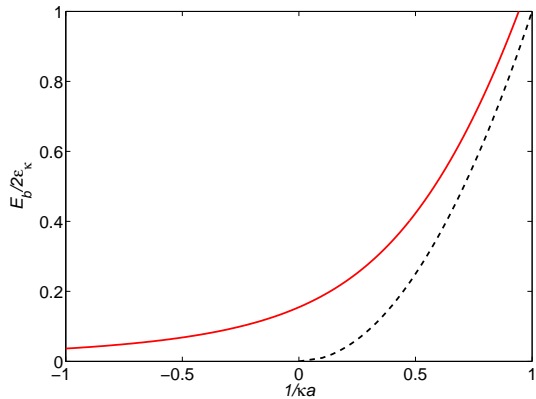}
\caption{\label{fig1}(color online) Binding energy of a diatomic molecule vs inverse of scattering length.
The solid line is the the binding energy with SOC, where the bound state exists for any scattering length and the binding energy vanishes at $a=0^{-}$.  The dashed line is the binding energy of a molecule without SOC which exist only with positive scattering length and vanishes as the scattering length diverge.}
\end{figure}

The binding energy of the bound state is defined as $E_{b}=-E_{0}-2\epsilon_{\kappa}$, where $\epsilon_{\kappa}=\hbar^2\kappa^2/2m$ is the lowest energy of a single atom with SOC.  In FIG.~\ref{fig1}, the binding energy is plotted against the inverse of the scattering length.  Since the relation between the binding energy and the corresponding scattering length is the same in all scattering channels, we drop the subscripts and denote the scattering length as $a$ in this plot.  As the scattering length decreases, the binding energy increases monotonously.  The binding energy vanishes when the scattering length $a$ approaches negative zero $0^-$, signaling that the resonance position is shifted from where $a$ diverges to $0^-$ and the bound state can exist with any value of $a$.  In the limit of $ a\rightarrow0^{-}$, we obtain the asymptotic form $E_{b}\sim \hbar^{2}\kappa^{4}a^2/(9m)$; at $1/a=0$, the binding energy is given by $E_{b}=(2\sqrt{3}-3)\hbar^{2}\kappa^{2}/(3m)$; when $\kappa a\rightarrow0^{+}$, the binding energy recovers the result in the case without SOC, $E_{b}\sim\hbar^{2}/(ma^{2})$.

The reason for the existence of the bound state for all values of the scattering length with the resonance position shifted to $0^{-}$ is the special single-particle density of states (DOS) due to SOC.  With SOC, the DOS at the lowest atom energy $\epsilon_\kappa=\hbar^2\kappa^2/2m$ is a constant, in sharp contrast to the case without SOC where DOS vanishes near the lowest atom energy.  As a result, the r.h.s. of Eq. (\ref{bound2}) has infrared divergence at $E_{0}=-2\epsilon_{\kappa}$ which guarantees a solution for any scattering length, whereas without SOC such infrared divergence is absent and the bound state only exists with positive scattering length.

The DOS effect on bound states due to SOC was first found in the case of two fermions \cite{PhysRevB.83.094515}.  The two-boson problem is more complicated due to statistics.  In the fermion case, the only s-wave interaction is the inter-species interaction, whereas in the boson case there are both inter-species and intra-species interactions.  In the case with two-bosons with Rashba SOC \cite{intro301}, due to statistics only the intra-species bound state is affected by SOC, while in the fermion case the inter-species bound state is affected.

\section{Pairing state of a Bose gas with SOC}
\subsection{Mean-field theory of pairing state}
In a Bose gas with an isotropic SOC, pairing of two atoms, i.e. the tendency of two atoms forming a diatomic molecule, may lead to the formation of molecular condensation at low temperatures.  This condensed state can be described by the BCS pairing theory.  To avoid the possibility of phase separation, here we consider the case with repulsive intra-species interactions and attractive inter-species interaction.  In this case the binding energy of the diatomic molecule in the inter-species channel is much smaller and this type of molecules are much easier to generate.  Thus we consider pairing between atoms with different spins only, with order parameter given by $\Delta=(g_{\uparrow\downarrow}/V)\sum_{\textbf{k}}\langle c_{-\textbf{k}\uparrow}c_{\textbf{k}\downarrow}\rangle$.  In general, the phase of the order parameter can be tuned arbitrarily under $U(1)$ gauge transformation, and in the following for simplicity we choose $\Delta>0$.

We study a spin-balanced Bose gas with an isotropic SOC at zero temperature in the mean-field approximation, where in addition to pairing the Hartree-Fock contributions are also included.  The mean-field Hamiltonian of this system is given by
\begin{equation}
H_{MF}=\frac{1}{2}\sum_{\textbf{k}}(B^{\dag}_{\textbf{k}}H_{\textbf{k}}B_{\textbf{k}}-2\xi_{\textbf{k}})
-\frac{\Delta^{2}}{g_{\uparrow\downarrow}}V-(2g_{\uparrow\uparrow}+g_{\uparrow\downarrow})n^{2}V, \label{M-F-H}
\end{equation}
where $\xi_{\textbf{k}}=\epsilon_{\textbf{k}}+2g_{\uparrow\uparrow}n+g_{\uparrow\downarrow}n-\mu$, $B^{\dag}_{\textbf{k}}$ is the field operator with four components $[c^{\dag}_{\textbf{k}\uparrow}, c_{-\textbf{k}\uparrow}, c^{\dag}_{\textbf{k}\downarrow}, c_{-\textbf{k}\downarrow}]$, $n$ is the atom density
of one spin component, and  $\mu$ is the chemical potential.  The matrix $H_{\textbf{k}}$ is given by
\begin{eqnarray}
H_{\textbf{k}}=\left[
\begin{matrix}
\xi_{\textbf{k}}+\frac{\hbar^{2}\kappa k_{z}}{m}&0&S^{*}(\textbf{k}_{\perp})&\Delta\\
0&\xi_{\textbf{k}}-\frac{\hbar^{2}\kappa k_{z}}{m}&\Delta&-S(\textbf{k}_{\perp})\\
S(\textbf{k}_{\perp})&\Delta&\xi_{\textbf{k}}-\frac{\hbar^{2}\kappa k_{z}}{m}&0\\
\Delta&-S^{*}(\textbf{k}_{\perp})&0&\xi_{\textbf{k}}+\frac{\hbar^{2}\kappa k_{z}}{m}
\end{matrix}
\right].\nonumber\\
\end{eqnarray}

The mean-field Hamiltonian Eq. (\ref{M-F-H}) can be diagonalized by the generalized Bogoliubov transformation.
We obtain two branches of quasi-particles with excitation energies given by
\begin{widetext}
\begin{equation}
\varepsilon_{\textbf{k}\pm}=\sqrt{\xi^{2}_{\textbf{k}}-\Delta^{2}+({\hbar k \kappa \over m})^2 \pm2 \frac{\hbar^{2}\kappa}{m}\sqrt{
k^2\xi^{2}_{\textbf{k}}-\Delta^{2}\textbf{k}^{2}_{\perp}}}.
\end{equation}
\end{widetext}
The energy gap, i.e. the smallest energy, of these excitations is given by $\varepsilon_{0}=\sqrt{\xi^{2}_{0}-\Delta^{2}}$.
For finite k, these excitation energies are isotropic in the $k_x$-$k_y$ plane, but anisotropic in the $k_x$-$k_z$ plane.
In the limit $\Delta\rightarrow0$, they recover the normal-state form, $\xi_{\textbf{k}}\pm {\hbar k \kappa / m}$.  For fixed $k$, the excitation energy of the lower branch $\varepsilon_{\textbf{k}-}$ is between $\sqrt{(\xi_{\textbf{k}}-{\hbar k \kappa / m})^2-\Delta^{2}}$ at $k_\perp=0$ and $\sqrt{\xi_{\textbf{k}}^2-\Delta^{2}}-{\hbar k \kappa / m}$ at $k_z=0$, while the excitation energy of the upper branch is between $\sqrt{(\xi_{\textbf{k}}+{\hbar k \kappa / m})^2-\Delta^{2}}$ and $\sqrt{\xi_{\textbf{k}}^2-\Delta^{2}}+{\hbar k \kappa / m}$.  The gap between the lower and the upper excitation branches is given by $2{\hbar k \kappa / m}$ at $k_z=0$.
%\begin{figure}
%\includegraphics[width=8.5cm]{fig4.eps}\label{excitation}
%\caption{\label{fig4}(color online)Anisotropy of the quasi-particle excitation energies $\varepsilon_{\textbf{k}\pm}$ at $\mu'/\epsilon_{\kappa}=-1,\Delta/\epsilon_{\kappa}=0.99$ and $k/\kappa=0.001$,where $\mu'=\mu-2g_{\uparrow\uparrow}n-g_{\uparrow\downarrow}$.}
%\end{figure}

The anisotropy of quasi-particle excitation energies is a consequence of spin-momentum locking due to SOC and pairing.  In the absence of pairing, the momentum and spin of a quasi-particle are locked, either parallel or anti-parallel due to SOC.  With pairing between spin-up and spin-down atoms, if the quasi-particle momentum is in $z$-direction, the spin-momentum locking is still present, and the excitation energy of the lower-branch quasi-particle is at minimum for fixed $k$; if the quasi-particle momentum is in $x$-$y$ plane, the spin-momentum locking is lost and the excitation energy of the lower-branch quasi-particle is at maximum.  The energy dependence of the upper-branch quasi-particle is simply opposite.

\begin{figure}
\includegraphics[width=8.5cm]{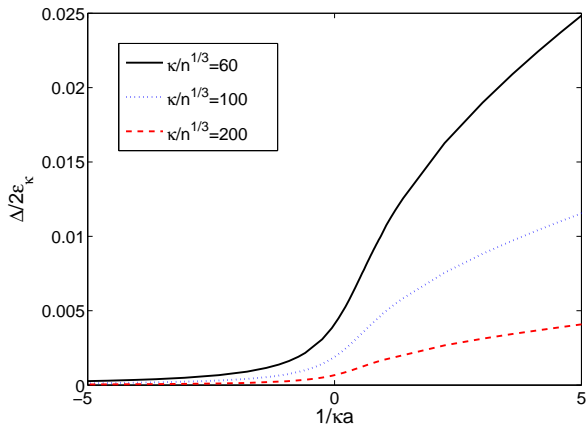}
\caption{\label{fig2}(color online) Pairing order parameter vs inverse of inter-species scattering length for several different densities.  In the dilute limit the solution of the order parameter always exists.}
\end{figure}

The pairing order parameter $\Delta$ and the chemical potential $\mu$ can be self-consistently solved together numerically.
We find that the mean-field solution always exists in the dilute limit $n \rightarrow 0$.  As shown in FIG.~\ref{fig2}, the order parameter increases monotonically with the inverse of the inter-species scattering length $1/a$.  In the limit $a \rightarrow 0^-$, the order parameter $\Delta$ vanishes; in the opposite limit $a \rightarrow 0^{+}$, $\Delta$ diverges.  The increase of the pairing order parameter with $1/a$ is consistent with relation between the binding energy $E_b$ of a diatomic molecule and $1/a$.
\\

%\begin{figure}
%\includegraphics[width=8.5cm]{fig5.eps}\label{mu}
%\caption{\label{fig5}(color online) Chemical potential vs particle densities at $1/\kappa a_{\uparrow\downarrow}=-4$ and $1/\kappa a_{\uparrow\uparrow}=0$.
%The chemical potential decrease
%with the density.}
%\end{figure}

\subsection{Phase transition}
The pairing state is always stable in the dilute limit with enough repulsive intra-species interaction.  As the density increases, the pairing order parameter increases, which reduces the energy gap of the quasi-particle excitation, contrary to the fermion case.  When the density increases to a critical value, the excitation gap vanishes.  Beyond the critical point, the pairing state do not exist and the system is likely turned into a BEC state of atoms.

In the BCS pairing state, the ground state energy consists of the kinetic, pairing, Hartree, and Fock energies.  The energy density of this pairing state is given by
\begin{eqnarray}\label{eg1}
E_{g1}={1 \over V}\sum_{\textbf{k}}[(\varepsilon_{\textbf{k}+}+\varepsilon_{\textbf{k}-})/2
-\xi_{\textbf{k}}+\Delta^{2}/(2\epsilon_{\textbf{k}})]-\nonumber\\
{ \Delta^{2} m \over 4\pi \hbar^2  a_{\uparrow\downarrow}}
+2\mu'n+ n^{2} {4\pi \hbar^2 \over m} (2a_{\uparrow\uparrow}+a_{\uparrow\downarrow}),
\end{eqnarray}
where $\mu'=\mu-(4 \pi \hbar^2/m)(2a_{\uparrow\uparrow}+a_{\uparrow\downarrow})n$.

In the atomic BEC state, the atoms condense into an equal-weight superposition state of two helical states with opposite spin in the lowest branch.  The energy density of this state is given by
\begin{equation}\label{eg2}
E_{g2}=-2 \epsilon_\kappa n+ {4\pi \hbar^2 \over m} n^{2}(a_{\uparrow\downarrow}+a_{\uparrow\uparrow}),
\end{equation}
where $n$ is the density of one spin component.

\begin{figure}
\includegraphics[width=8.5cm]{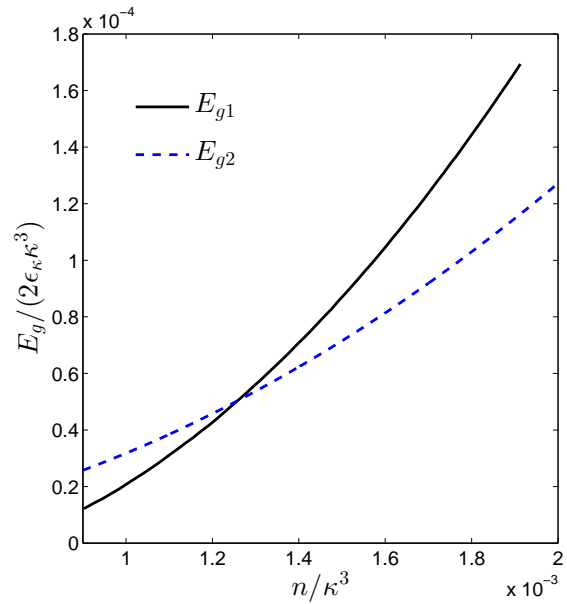}
\caption{\label{fig3}(color online) Energy comparison between atomic and molecular condensates at $\kappa a_{\uparrow\downarrow}=-1.49$ and $\kappa a_{\uparrow\uparrow}=4.02$.  The solid line is the energy density of the BCS state, $E_{g1}$, and the dashed line is the energy density of the atomic BEC state, $E_{g2}$.  Both are subtracted by $-2\epsilon_\kappa n$.  The transition from BCS to BEC takes place around $n/\kappa^3\approx1.26\times10^{-3}$.}
\end{figure}

Comparing the two energies, Eq. (\ref{eg1}) and (\ref{eg2}), we see that in the BCS pairing state the order parameter $\Delta$ reduces the total energy, but the Hartree energy in the pairing state is twice of that in the atomic BEC state.  Therefore as the density increases, there is transition from the BCS pairing state to the atomic BEC state.  From numerical calculation, we find that when the excitation gap of the pairing state vanishes, the atomic BEC state energy is smaller than the pairing state energy, $E_{g1}>E_{g2}$, indicating that even before that gap closes, the system has already turned into the atomic BEC state and this transition is a first-order phase transition as shown in Fig. \ref{fig3}.\\

\section{Discussion and Conclusion}.
In our previous work on a Bose gas with Rashba SOC \cite{intro301}, we considered the case with intra-species attraction and inter-species repulsion, where the pairing state may suffer phase separation.  In this work, a Bose gas with spherical SOC is studied with intra-species repulsion and inter-species attraction, which can avoid phase separation.  The two-body bound state structures are different between Rashba and spherical SOC cases.  In the Rashba SOC case, the resonance position of the intra-species scattering is shifted, while there is no shift for inter-species scattering.  In the spherical SOC case, the resonance positions for all scatterings are shifted.  With intra-species repulsion and inter-species attraction, the binding energy of the intra-species molecule is much larger than that of the inter-species molecule.  Therefore it will be much easier to generate the inter-species pairing state.

In the case with Rashba SOC, a superfragmented state with very large degeneracy was found when two atoms are restricted to the single-particle ground states \cite{PhysRevLett.110.140407}.  In our work with spherical SOC, we did not make any approximation and found three bound states, two from intra-species scatterings and one from inter-species scattering.  The pairing state that we are studying is a dilute BEC state of diatomic molecules. It is a symmetry-breaking state with off-diagonal long range order, and should be stable against ordinary perturbations such as anisotropy of SOC. When SOC is anisotropic, as long as its dimensionality and the lower-energy behavior of the single-particle density of states are unchanged, there is no qualitative change in either the two-body bound state structure or the pairing state.

A big obstacle in creating the BCS pairing state of bosons is the particle loss due to three-body recombination near the resonance.  For a Bose gas with SOC, two-body scattering has been theoretically investigated \cite{PhysRevA.86.053608,PhysRevA.87.052708}, but little is quantitatively known about three-body scattering.  Qualitatively, the relevance of the three-body recombination is indicated by the parameter $n_{a}b^3$, where $n_{a}$ is the atom density and $b$ is the size of the diatomic molecule inversely proportional to the square root of binding energy, $b=\hbar/\sqrt{mE_b}$.  If $n_{a}b^3 \ll 1$, the three-body recombination is not dominant.  We find that under current experimental conditions a stable pairing state can indeed exist.  For an ultra-cold Bose gas with atom density about $n_a=2 \times 10^{17}m^{-3}$ and the isotropic SOC $\kappa=2.51\times10^{7}m^{-1}$, a BCS paring state can exist with the inter-species scattering length $a_{\uparrow\downarrow}=-59.3nm$ and the intra-species scattering length $a_{\uparrow\uparrow}=160nm$ which corresponds to $b_{\uparrow\downarrow}=171nm$ and $b_{\uparrow\uparrow}=79nm$ with $n_{a}b_{\uparrow\downarrow}^3=0.001$ and $n_{a}b_{\uparrow\uparrow}^3=0.0001$.

Beyond mean-field approximation, there are important pairing fluctuations \cite{PhysRevA.79.063609}.  In the parameter region that we are considering, the Bose gas is dilute, and at low temperatures it is simply a weakly-interacting gas of diatomic molecules.  The critical temperature can be estimated by the BEC temperature of an ideal Bose gas.  Since the effective mass of the molecule is of the order of twice the atom mass and the molecule density is half the atom density, the critical temperature is of the order of a quarter of ideal atom BEC temperature.  We plan to study the fluctuation effect in our future works.

In summary, we study a two-component Bose gas with a spherical SOC and find that two atoms can form a bound state with any intra- or inter-species scattering lengths due to the SOC effect on DOS.  In the dilute limit, a stable BCS pairing state can be formed with attractive inter-species and repulsive intra-species interactions.  The excitation energies of the pairing state are anisotropic.  As the density increases, there is a first-order transtition from the BCS to BEC states.

\begin{acknowledgements} We would like to thank Z.-Q. Yu, R. Li, P. Zhang, L. You, and T.-L. Ho for helpful discussions.  This work is supported by NSFC under Grant No 11274022 and NKRDP under Grant No. 2016YFA0301500.
\end{acknowledgements}

\end{document}